\documentclass[aps,prb,superscriptaddress,reprint, amsmath]{revtex4-1}
\usepackage{hyperref}
\usepackage{graphicx, color}
\usepackage[normalem]{ulem}
\newcommand{\be}{\begin{equation}}
\newcommand{\ee}{\end{equation}}

\newcommand{\bra}[1]{\langle #1|}
\newcommand{\ket}[1]{|#1 \rangle}

\newcommand{\EnL}{\varepsilon_L}
\newcommand{\EnR}{\varepsilon_R}
\newcommand{\EnM}{\varepsilon_M}

\newcommand{\EM}{E_\text{M}}

\newcommand{\DQDdetn}{\delta}

\newcommand{\gstate}{\ket{0}}
\newcommand{\estate}{\ket{1}}
\newcommand{\gsDQD}{\ket{0_\text{DQD}}}
\newcommand{\esDQD}{\ket{1_\text{DQD}}}
\newcommand{\braestate}{\bra{1}}

\newcommand{\symstate}{\ket{\mathcal{S}}}
\newcommand{\leakstate}{\ket{\mathcal{L}}}
\newcommand{\braleakstate}{\bra{\mathcal{L}}}
\newcommand{\leftstate}{\ket{L}}
\newcommand{\midstate}{\ket{M}}
\newcommand{\rightstate}{\ket{R}}

\newcommand{\EQ}{E_\text{Q}}

\newcommand{\Egei}{\varepsilon_1}
\newcommand{\Egeii}{\varepsilon_2}
\newcommand{\Egeiii}{\varepsilon_3}

\newcommand{\VtL}{V_\text{t,L}}
\newcommand{\VtR}{V_\text{t,R}}
\newcommand{\VdL}{V_\text{L}}
\newcommand{\VdM}{V_\text{M}}
\newcommand{\VdR}{V_\text{R}}

\newcommand{\fd}{\nu_\text{d}}
\newcommand{\fr}{\nu_\text{r}}
\newcommand{\fri}{\nu_\text{r,1}}
\newcommand{\frii}{\nu_\text{r,2}}
\newcommand{\fs}{\nu_\text{s}}
\newcommand{\fp}{\nu_\text{p}}

\newcommand{\kint}{\kappa_\text{int} / 2\pi}
\newcommand{\kext}{\kappa_\text{ext} / 2\pi}

\newcommand{\figa}{\textbf{a}}
\newcommand{\figb}{\textbf{b}}
\newcommand{\figc}{\textbf{c}}
\newcommand{\figd}{\textbf{d}}
\newcommand{\fige}{\textbf{e}}
\newcommand{\figf}{\textbf{f}}

\newcommand{\figta}{a}
\newcommand{\figtb}{b}
\newcommand{\figtc}{c}
\newcommand{\figtd}{d}
\newcommand{\figte}{e}
\newcommand{\figtf}{f}

\newcommand{\detuning}{\delta}

\newcommand{\Sii}{S_{11}}
\newcommand{\HWHM}{\delta\nu_\text{q}}

\newcommand{\CxA}{A}
\newcommand{\CxAo}{A_0}

\begin{document}

%%%
%%% TITLE
%%%

\title{Strong photon coupling to the quadrupole moment of an electron in solid state}

%%%
%%% AUTHORS
%%%

\author{J. V. Koski}
\affiliation{Department of Physics, ETH Zurich, CH-8093 Zurich, Switzerland}
\author{A. J. Landig}
\affiliation{Department of Physics, ETH Zurich, CH-8093 Zurich, Switzerland}
\author{M. Russ}
\affiliation{Department of Physics, University of Konstanz, D-78457 Konstanz, Germany}
\author{J. C. Abadillo-Uriel}
\affiliation{Department of Physics, University of Wisconsin-Madison, Madison, Wisconsin 53706, USA}
\author{P. Scarlino}
\affiliation{Department of Physics, ETH Zurich, CH-8093 Zurich, Switzerland}
\author{B. Kratochwil}
\affiliation{Department of Physics, ETH Zurich, CH-8093 Zurich, Switzerland}
\author{C. Reichl}
\affiliation{Department of Physics, ETH Zurich, CH-8093 Zurich, Switzerland}
\author{W. Wegscheider}
\affiliation{Department of Physics, ETH Zurich, CH-8093 Zurich, Switzerland}
\author{Guido Burkard}
\affiliation{Department of Physics, University of Konstanz, D-78457 Konstanz, Germany}
\author{Mark Friesen}
\affiliation{Department of Physics, University of Wisconsin-Madison, Madison, Wisconsin 53706, USA}
\author{S. N. Coppersmith}
\email{Present address: School of Physics, University of New South Wales, Sydney NSW 2052, Australia}
\affiliation{Department of Physics, University of Wisconsin-Madison, Madison, Wisconsin 53706, USA}
\author{A. Wallraff}
\affiliation{Department of Physics, ETH Zurich, CH-8093 Zurich, Switzerland}
\author{K. Ensslin}
\affiliation{Department of Physics, ETH Zurich, CH-8093 Zurich, Switzerland}
\author{T. Ihn}
\affiliation{Department of Physics, ETH Zurich, CH-8093 Zurich, Switzerland}

\date{\today}

\pacs{}
\maketitle

%%%
%%% ABSTRACT
%%%

\textbf{
The implementation of circuit quantum electrodynamics \cite{Wallraff2004} allows coupling distant qubits by microwave photons hosted in on-chip superconducting resonators \cite{Majer2007, DiCarlo2009, VanWoerkom2018}. 
Typically, the qubit-photon interaction is realized by coupling the photons to the electric dipole moment of the qubit. 
A recent proposal \cite{Friesen2017, Ghosh2017} suggests storing the quantum information in the electric quadrupole moment of an electron in a triple quantum dot. 
The qubit is expected to have improved coherence since it is insensitive to dipolar noise produced by distant voltage fluctuators. 
Here we experimentally realize a quadrupole qubit in a linear array of three quantum dots in a GaAs/AlGaAs heterostructure. 
A high impedance microwave resonator coupled to the middle dot interacts with the qubit quadrupole moment. 
We demonstrate strong quadrupole qubit--photon coupling and observe improved coherence properties when operating the qubit in the parameter space where the dipole coupling vanishes.
}

%%%
%%% FIGURE: 1
%%%

\begin{figure}[h!t]
\includegraphics[width=\columnwidth]{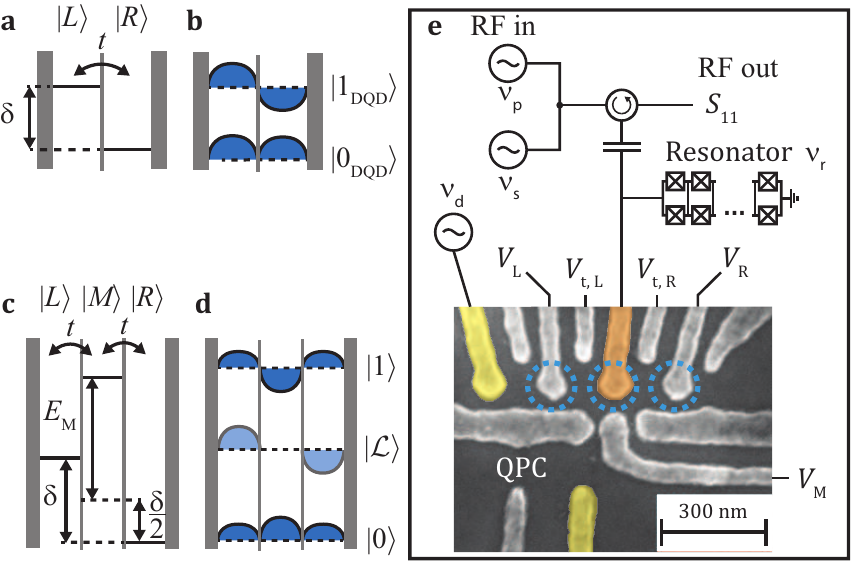}
\caption{\textbf{Schematics of the qubit wavefunctions and the device.} 
\figa, Energy level diagram of a double quantum dot (DQD). 
The states $\leftstate$ and $\rightstate$ are detuned by $\detuning$ and tunnel coupled by $t$.
\figb, Wavefunctions $\ket{\psi} = \frac{1}{\sqrt{2}}\leftstate \pm \frac{1}{\sqrt{2}}\rightstate$ of the ground $\gsDQD$ and excited $\esDQD$ states of a DQD charge qubit for $\DQDdetn = 0$. 
\figc, Energy level diagram of a triple quantum dot (TQD). 
\figd, Wavefunctions $\ket{\psi}$ of the ground $\gstate$, leakage $\leakstate$, and excited $\estate$ states of a quadrupole qubit in a TQD for $\EM = 0, \detuning = 0$. 
\fige, Schematic of the measurement setup and a scanning electron micrograph image of the TQD device. 
Dashed circles indicate the positions of the quantum dots. 
The number of electrons in the quantum dots is monitored with a quantum point contact (QPC). 
Voltages $\VdL$, $\VtL$, $\VtR$, $\VdR$, and $\VdM$ control the quantum dot energy levels and the interdot tunnel couplings. 
The gates highlighted in color are at DC ground. The gate marked in orange is directly connected to an on-chip high impedance quarter wavelength resonator with a tunable resonance frequency $\fr$. The resonator consists of an array of pairwise parallel Josephson junctions (black rectangles with crosses) \cite{Stockklauser2017}.
The resonator reflection coefficent $S_{11}$ is probed at frequency $\fp$. 
The qubit can be driven either via the resonator at frequency $\fs$, or directly via the indicated gate at frequency $\fd$. 
\label{Fig1}} 
\end{figure}

%%%
%%% PARAGRAPH: INTRODUCTION, DOUBLE QUANTUM DOTS
%%%

A single electron confined in two tunnel-coupled quantum dots forms a double quantum dot (DQD) charge qubit \cite{Frey2012}, where the logical qubit states $\gsDQD$ and $\esDQD$ are spanned by the occupation states $\leftstate$ and $\rightstate$ of the electron localized in the left dot at electrochemical potential $\EnL$ or in the right dot at potential $\EnR$. 
As illustrated in Fig.\ \ref{Fig1} \figta-\figtb, the electron wavefunctions hybridize when the tunnel coupling $t$ is high compared to the detuning $\DQDdetn = \EnL - \EnR$, introducing a finite dipole moment $p \propto -ed$, where $-e$ is the electron charge and $d$ is the distance between the two quantum dots, to the qubit transition. 
The qubit can then be coupled to single microwave photons by engineering the vacuum voltage fluctuations $V_0$ of the microwave resonator to modulate $\DQDdetn$, resulting in a coupling strength $g \propto pV_0$. 
This way, strong coupling between a microwave photon and a DQD charge qubit has been realized experimentally \cite{Stockklauser2017, Mi2017, Bruhat2018}. 
Similarly, strong coupling of microwave photons to spin qubits has been achieved by introducing an electrical dipole moment to the spin degree of freedom \cite{Viennot2015, Mi2018, Samkharadze2018, Landig2018}.
While the susceptibility to $\DQDdetn$ is essential for the qubit-photon coupling, the qubit energy dependence on $\DQDdetn$ leads to dephasing by charge noise\cite{Bruhat2018}. 
The noise-induced dephasing is minimal where the qubit energy dispersion has a zero first-order derivative. 
Such flat dispersions are often referred to as `sweet spots' \cite{Vion2002, Thorgrimsson2017, Zajac2018} and are the preferred qubit operation points.
However, practical qubit pulsing schemes may require the qubit to be driven out of the sweet spot \cite{Yoneda2018, Watson2018}, subjecting it to increased decoherence during the pulses. 

%%%
%%% PARAGRAPH: INTRODUCTION TO THE QUADRUPOLE QUBIT
%%%

Here, we experimentally investigate the charge quadrupole qubit proposed in \onlinecite{Friesen2017} that can be manipulated while constantly maintaining the qubit in a sweet spot. 
The proposed qubit is formed by a single electron confined in a triple quantum dot (TQD). 
The qubit is spanned by the occupation states $\leftstate$, $\midstate$, and $\rightstate$, shown in Fig.\ \ref{Fig1}\figtc, of an electron localized in the left, the middle, or the right dot, respectively. 
The electrochemical potentials $\EnL$, $\EnM$, and $\EnR$ of these states are parametrized by the left-right detuning $\detuning = \EnL - \EnR$ and the middle-dot detuning $\EM = \EnM - (\EnL + \EnR) / 2$. 
Furthermore, the nearest-neighbor quantum dots are tunnel coupled with equal tunneling amplitudes $t$. 
The system has three eigenstates of which the logical qubit states are the one with the lowest eigenenergy $\Egei$ and the one with the highest eigenenergy $\Egeiii$.
This introduces a `leakage state' with energy $\Egeii$ that lies energetically between the qubit states.
The quadrupole qubit is realized at zero detuning $\detuning = 0$ where for any $\EM$ the qubit energy $\EQ = \Egeiii - \Egei$ has a sweet spot in $\detuning$, although not in $\EM$.
Since the quadrupolar detuning $\EM$ has a higher order multipole character than the dipolar detuning $\detuning$, the noise, characterized by standard deviations $\sigma_\detuning$ and $\sigma_{\EM}$ and caused by a noise source at a distance $R$ from the qubit, is expected\cite{Friesen2017} to follow $\sigma_\detuning / \sigma_{\EM} \propto d / R$. This implies that if the noise sources are in general distant from the triple dot, $R \gg d$, the qubit retains a comparably long coherence time even if $\EM$ is detuned. Conversely, determining the ratio $\sigma_\detuning / \sigma_{\EM}$ provides spatial information about the origin of noise. This is in contrast to standard DQD charge qubits that have only one degree of freedom $\detuning$ to which electric noise can couple (typically, the voltage noise sensitivity of $t$ is orders of magnitude weaker).

%%%
%%% PARAGRAPH: DEFINE THE QUADRUPOLE QUBIT
%%%

The quadrupole qubit is realized at zero detuning $\delta = 0$, where the logical qubit states are $\gstate =\cos(\phi/ 2) \symstate + \sin(\phi/ 2) \midstate$ and $\estate =\sin(\phi/ 2) \symstate - \cos(\phi/ 2) \midstate$ with $\symstate = \frac{1}{\sqrt{2}}(\leftstate + \rightstate)$, the mixing angle $\phi$ is determined by $\cos(\phi) = \EM / \EQ$, and $\EQ = \sqrt{8t^2 + \EM^2}$. The qubit states as well as the leakage state $\leakstate = \frac{1}{\sqrt{2}}\left(\leftstate - \rightstate\right)$ are illustrated for $\detuning = \EM = 0$ in Fig.\  \ref{Fig1}\figtd.
Due to the symmetry of the wavefunctions $\gstate$ and $\estate$, the qubit has a small dipole moment characterized by the matrix element $p_x = -e |\braestate x\gstate|$ which is zero in the ideal limit where the three quantum dots are uniaxial and equidistant, implying that coupling to a microwave photon cannot be realized via $\detuning$.
However, the qubit quadrupole moment characterized by the matrix element $Q_{xx} = -2 e |\braestate x^2\gstate|$ ($= -ed^2\sin(\phi)$ in the ideal limit) is finite, facilitating a finite qubit-photon coupling via $\EM$.
Conversely, the leakage state transition quadrupole matrix elements $Q^{\mathcal{L},0}_{xx} = -2 e |\braleakstate x^2\gstate|$ and $Q^{\mathcal{L},1}_{xx} =  -2 e |\braleakstate x^2\estate|$ are small due to the asymmetry of the leakage state $\leakstate$, and zero for a perfectly symmetric triple dot.
Therefore, by driving the qubit only via $\EM$, drive-induced transitions to the leakage state are strongly suppressed. 
The properties and potential benefits of qubits with quadrupolar character have been further studied in \cite{Oi2005, Ghosh2017, Kornich2018, Russ2018}.
We discuss the role of spin and magnetic field noise in the Supplementary Information \cite{Supplementary} and note that they are similar to conventional charge qubits in that the qubit energetics are equivalent for a spin-up and a spin-down electron.

%%%
%%% FIGURE: 2
%%%

\begin{figure}[h!b]
\includegraphics[width=\columnwidth]{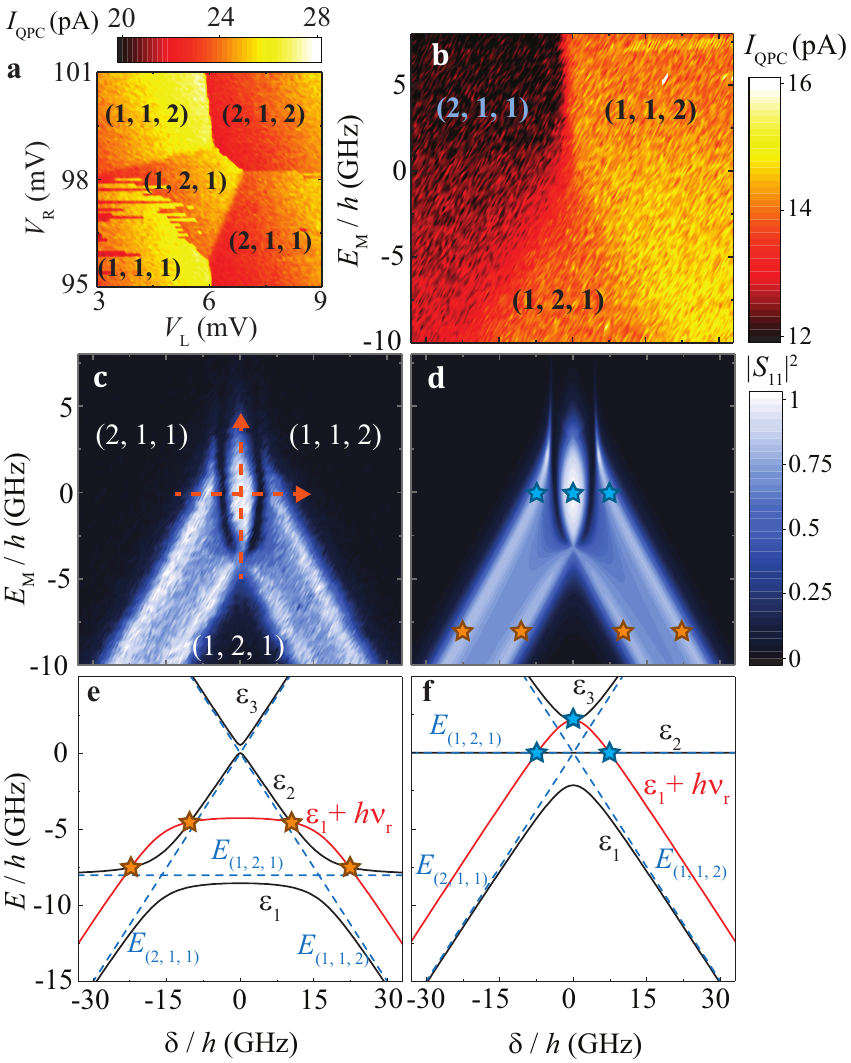}
\caption{\textbf{Qubit charge configuration and resonator amplitude response.} 
\figa, The current $I_\text{QPC}$ through the QPC as a function of $\VdL$ and $\VdR$. 
($L, M, R$) denotes $L$ electrons in the left, $M$ in the middle, and $R$ in the right dot.
\figb, $I_\text{QPC}$ as a function of $\EM$ and $\delta$, both of which are tuned as a linear combination of $\VdL$, $\VdM$, and $\VdR$. 
The relevant charge states for a quadrupole qubit, $(2, 1, 1)$, $(1, 2, 1)$, and $(1, 1, 2)$, are discernible in the data. 
\figc, Normalized reflected probe tone amplitude $|\Sii|^2$ at $\fp = \fri$ in the same parameter window as panel \figb.
The orange dashed lines indicate the measurement axes in Figs.\ \ref{Fig3} and \ref{Fig4}. 
\figd, $|\Sii|^2$ as obtained from an input-output model described in Supplementary Information\cite{Supplementary}.
\fige, The TQD energy level diagram at $\EM / h = -8$ GHz as a function of $\detuning$. \figf, the level diagram at $\EM / h = 0$ GHz. The dashed blue lines show the chemical potentials of the three relevant charge states and the solid black lines show the TQD eigenenergies. The red line shows the ground state energy offset by the resonator photon energy $h\fri = h \times 4.269$ GHz. At the positions marked by yellow or blue stars a resonance condition $\varepsilon_{2/3} - \varepsilon_1 = h\fri$ is satisfied. The corresponding detuning values are indicated in panel \figd.
\label{Fig2}}
\end{figure}

%%%
%%% PARAGRAPH: DEVICE CALIBRATION AND FIRST MEASUREMENTS
%%%

Our experimental implementation of the qubit is shown in Fig.~\ref{Fig1}\figte. 
Three quantum dots are formed in a GaAs/AlGaAs heterostructure by electrostatic confinement with aluminum top gate electrodes.  
The tunnel couplings between the quantum dots are controlled by voltages $\VtL$ and $\VtR$, while the dot electrochemical potentials are controlled by $\VdL$, $\VdM$, and $\VdR$. 
A high impedance frequency-tunable quarter wavelength resonator \cite{Stockklauser2017} is directly connected to the plunger gate overlapping the middle dot. There, the resonator vacuum voltage fluctuations modulate $\EM$, realizing quadrupolar coupling. 
Throughout the experiments we estimate the average population of photons in the cavity to be less than one.
A quantum point contact (QPC in Fig.\ \ref{Fig1}\figte) near the quantum dots acts as a sensor for the number of electrons in the quantum dot system, as shown in Fig.\ \ref{Fig2}\figta. 
We form the quadrupole qubit with the charge states $(1, 2, 1), (1, 1, 2)$ and $(2, 1, 1)$, where (L, M, R) denotes the number of electrons in the left, middle, and right dot. 
We show in the Supplementary Information \cite{Supplementary} that this configuration realizes the quadrupole qubit Hamiltonian.
We tune $\detuning$ and $\EM$ by\cite{Supplementary} changing a linear combination of $\VdL$, $\VdM$, and $\VdR$ that keeps the charge states $(1, 1, 1)$ and $(2, 1, 2)$ higher in energy, allowing us to map the full parameter space of the quadrupole qubit as shown in Fig.~\ref{Fig2}\figtb. 
We set\cite{Supplementary} the interdot tunnel couplings to $t / h \simeq 1.51$ GHz by tuning the tunnel gates $\VtL$ and $\VtR$. 
This corresponds to a quadrupole qubit energy $\EQ / h = \sqrt{8} t / h\simeq 4.27$ GHz at $\detuning = \EM = 0$. 

%%%
%%%  PARAGRAPH: AMPLITUDE RESPONSE
%%%

Figure \ref{Fig2}\figtc~shows the measured normalized resonator reflection $|\Sii|^2$ probed at frequency $\fp = \fri$ as a function of the detuning parameters $\detuning$ and $\EM$ in the quadrupole qubit regime. 
Here, the bare resonator frequency is set to $\fri = 4.269$ GHz with internal and external photon decay rates of $\kint \simeq 10$ MHz and $\kext \simeq 8$ MHz, respectively. 
The reflected signal reaches its minimum (black in Fig.\ \ref{Fig2}\figtc-\figtd) when the resonator photons do not interact with the TQD. 
Interaction with the TQD shifts the resonance frequency, causing $\fp$ to no longer probe on resonance. This increases $|\Sii|^2$ towards the background level (white in Fig.\ \ref{Fig2}\figtc-\figtd).
The signal expected by an input-output model\cite{Supplementary}, shown in Fig.\ \ref{Fig2}\figtd, is in good agreement with our experimental observations.
Away from the triple point, $|\detuning / h| > 5$ GHz, $\EM / h < -5$ GHz, at the cross-over between the $(1, 2, 1)$ and the $(2, 1, 1)$, or between the $(1, 2, 1)$ and the $(1, 1, 2)$ charge regimes, the system operates as a conventional DQD charge qubit between the left and the middle or the right and the middle quantum dots, respectively. 
As illustrated in the level diagram for $\EM / h = -8$ GHz shown in Fig.\ \ref{Fig2}\figte, there are two resonances $\varepsilon_2 - \varepsilon_1 = h\fri$ for both DQD configurations. These  appear as elevated reflection $|\Sii|^2$ at the resonance positions, observed as two lines of increased reflection in Figs.\ \ref{Fig2}\figtc-\figtd. In-between the resonances, the increased reflection is due to dispersive triple dot - resonator interaction. Centered at the triple point $\detuning = \EM = 0$, we find a signature of photon interaction with the quadrupole qubit due to the resonance $\varepsilon_3 - \varepsilon_1 = h\fri$, as shown in Fig. \ref{Fig2}\figtf. Close to the triple point, the transition energy can be approximated as $\varepsilon_3 - \varepsilon_1 \approx \sqrt{8t^2 + \detuning^2 + \EM^2}$, implying that increasing $|\detuning|$ or $|\EM|$ increases the qubit energy, detuning it from the resonator energy $h\fri$. This is apparent as a maximum in $|\Sii|^2$ at the triple point. Finally, around the triple point we find two arc-shaped reflection minima at approximately $\detuning / h = \pm 4$ GHz. At these minima, the dispersive shift from the transition corresponding to $\varepsilon_2 - \varepsilon_1$ and $\varepsilon_3 - \varepsilon_1$ have equal magnitude but opposite sign. Consequently, the resonance frequency is unchanged and therefore coincides with $\fp$.

%%%
%%% FIGURE:3
%%%

\begin{figure} [h!t]
\includegraphics[width=\columnwidth]{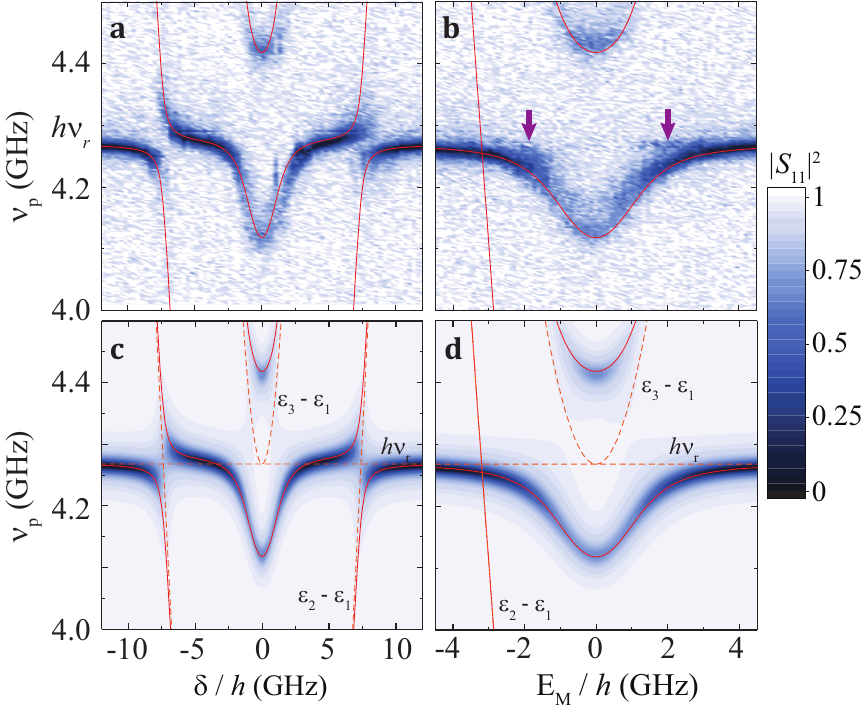}
\caption{\textbf{Qubit-photon interaction.} 
\figa, Measured and \figc, simulated \cite{Supplementary} reflection $|S_{11}|^2$ as a function of $\fp$ and $\delta$ at $\EM \simeq 0$. 
\figb, Measured and \figd, simulated \cite{Supplementary} reflection $|S_{11}|^2$ as a function of $\fp$ and $\EM$ at $\delta \simeq 0$.
The solid red lines show the expected eigenenergies of the hybrid quantum dot-resonator system. 
The dashed orange lines (in panels \figc-\figd) show the individual dispersions of the triple quantum dot and the resonator.
The purple arrows mark a finite signal at the bare resonance frequency arising from a finite population of the leakage state.
\label{Fig3}}
\end{figure}

%%%
%%% PARAGRAPH: PHOTON INTERACTION WITH A QUADRUPOLE QUBIT
%%%

Next, we investigate the resonant interaction between a microwave photon and the quadrupole qubit as a function the dipolar detuning $\delta$ with $\EM = 0$, and the quadrupolar detuning $\EM$ with $\delta = 0$. 
We measure the normalized resonator reflection $|\Sii|^2$ as a function of resonator probe frequency $\fp$ in the corresponding detuning range for $\detuning$ as shown in Fig.~\ref{Fig3}\figta, which is simulated with an input-output model in Fig.~\ref{Fig3}\figtc. 
Here, we observe three avoided crossings in the energy dispersion of the hybrid quantum dot-photon system.
The two avoided crossings at $\detuning / h \simeq \pm 7$ GHz originate from the interaction between the resonator photon and the transition between the lowest and the second lowest energy eigenstates (qubit ground state and leakage state) with energies $\varepsilon_1$ and $\varepsilon_2$ as shown in Fig.\ \ref{Fig3}\figc. Note that here these two transitions have a finite quadrupole moment and therefore a finite coupling to the resonator since $\detuning$ is non-zero.
The third observed avoided crossing centered at $\detuning = 0$ is due to the interaction between the resonator photon and the transition between the lowest and the highest energy eigenstates with energies $\varepsilon_1$ and $\varepsilon_3$ (qubit ground and excited state).

%%%
%%% FIGURE:4
%%%

\begin{figure} [h!!t]
\includegraphics[width=\columnwidth]{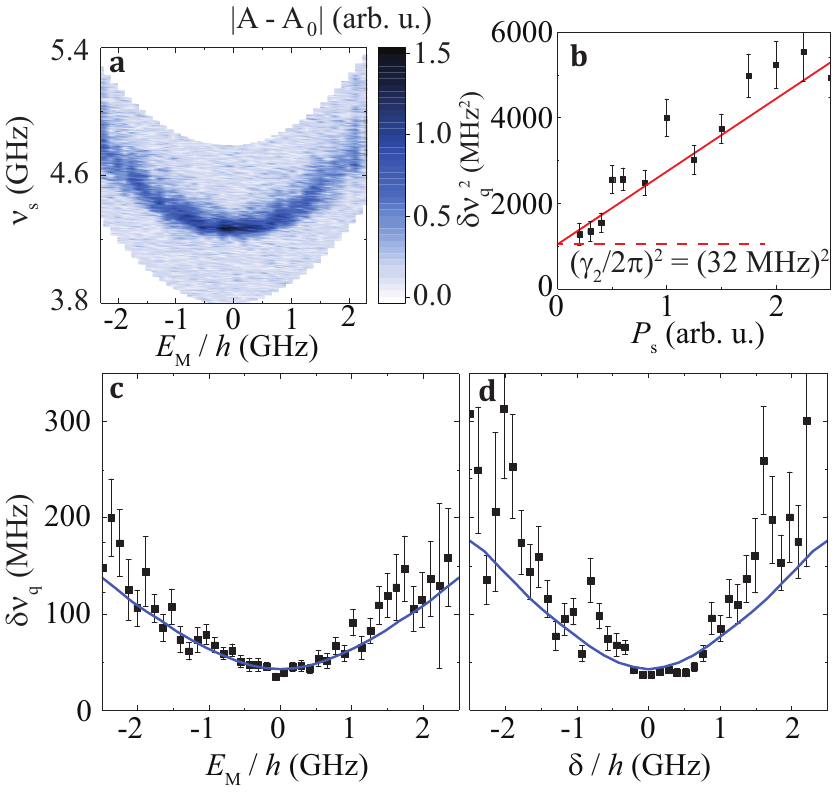}
\caption{\textbf{Determination of qubit energy and linewidth}. 
\figa, The complex amplitude change $|\CxA - \CxAo|$ of the reflected signal as a function of spectroscopy frequency $\fs$ and detuning $\EM$. 
\figb, Half-width-half-maximum $\HWHM^2$ of the resonances in $|\CxA - \CxAo|$ at $\detuning = \EM = 0$ as a function of drive power $P_\text{s}$. 
The red line shows a linear fit to the data points. 
The red dashed line corresponds to the qubit decoherence $\gamma_2$, which is determined from the extrapolated $\HWHM$ at zero drive power. 
\figc, Measured qubit linewidth as a function of $\EM$ at $\detuning = 0$. 
\figd, Measured qubit linewidth as a function of $\detuning$ at $\EM = 0$.
The solid lines in \figc~and \figd~are theoretical fits to a model of electric noise-induced dephasing (see Supplementary Information \cite{Supplementary}).
\label{Fig4}}
\end{figure}

%%%
%%% PARAGRAPH: PHOTON INTERACTION WITH A QUADRUPOLE QUBIT II
%%%

Figures \ref{Fig3}\figtb~and \ref{Fig3}\figtd~present the measured and the simulated\cite{Supplementary} normalized resonator reflection $|\Sii|^2$ as a function of $\EM$ and $\fp$ while $\detuning = 0$. 
In this configuration, the system eigenstates are the quadrupole qubit states $\gstate$ and $\estate$ and the leakage state $\leakstate$.
We observe strong quadrupole qubit - photon interaction with an estimated qubit-photon coupling strength $g_0 / 2\pi \simeq 150$ MHz determined from the vacuum Rabi mode splitting. 
Conversely, we observe that the coupling to the leakage state transition is negligible, apparent in the measured and simulated data as an absence of avoided crossing at the resonator photon - leakage state transition resonance.
Instead, we observe a finite decrease in reflection at the bare resonance frequency $\fri = 4.269$ GHz superposed with the signal from qubit-photon interaction at $\detuning = 0$ for $\EM$ values indicated in Fig.~\ref{Fig3}\figtb.
We interpret this as a signature of a finite leakage state population due to slow incoherent phonon- \cite{Gasser2009} or photon-induced transitions from the logical qubit states.

%%%
%%% PARAGRAPH: LINEWIDTH MEASUREMENTS
%%% 

To perform quadrupole qubit spectroscopy using the resonator for dispersive readout\cite{Schuster2005}, we set the resonator frequency to $\frii \simeq 6.215$ GHz such that the qubit and the resonator are detuned by $h\fr - \EQ > 10 \hbar g_0$. 
When set to $\frii$, the internal and external photon decay rates are $\kint = 6$ MHz and $\kext = 4$ MHz, respectively.
Here, the dispersive interaction between the qubit and the resonator shifts the resonator resonance frequency \cite{Blais2004} to $\tilde \fr \approx \frii - g^2 / (\frii - \EQ)$, where $g = 2g_0 c_M^0 c_M^1$ is the $\detuning$- and $\EM$-dependent qubit-photon coupling strength, and $c_M^{0(1)}$ are the wavefunction coefficents of the middle dot occupation of the state $\gstate$ ($\estate$).
We apply a probe tone at the shifted frequency $\fp \simeq \tilde \fr$ and measure the reflected complex amplitude $\CxA = I + iQ$ consisting of the in-phase $I$ and the quadrature-phase $Q$ components. 
Furthermore, we apply a spectroscopy tone $\fs$ that increases the qubit excited state population, if it is on resonance with the qubit transition frequency, $h\fs \simeq \EQ$. 
This is discernible in the reflected signal as the resonance frequency tends to the bare one, $\frii$. 
As shown in Fig.\ \ref{Fig4}\figta, this frequency shift is detected as a change in $\CxA$ and produces a resonance peak centered at $h\fs \simeq \EQ$ in $|\CxA - \CxAo|$, where $\CxAo$ is the signal amplitude in the absence of the qubit spectroscopy drive. 
The half-width-half-maximum $\delta\nu_q$ of the resonance peak yields the qubit decoherence rate $\gamma_2$ as $\delta\nu_q = \sqrt{(\gamma_2 / 2\pi)^2 + \beta P_\text{s}}$, where $\beta P_\text{s}$ is the drive power $P_\text{s}$ broadening of the resonance and $\beta$ is a constant \cite{Schuster2005}.
Figure \ref{Fig4}\figtb~shows the measured $\HWHM^2$ at $\detuning = \EM = 0$ as a function of $P_\text{s}$. 
We perform a linear fit to the data from which we determine the qubit decoherence rate $\gamma_2 / 2\pi \simeq 32$ MHz from the intersection point at zero drive power.
The charge decoherence rate is comparable to other implementations in GaAs charge qubits with few electrons \cite{Stockklauser2017, Landig2018}, however, not as low as $\gamma_2 / 2\pi \simeq 3$ MHz reported for silicon \cite{Mi2017} or for large quantum dots in GaAs \cite{VanWoerkom2018}. Further improvement in coherence can be reached by optimizing the electrostatic design of the quantum dots.

%%%
%%% PARAGRAPH: LINEWIDTH MEASUREMENTS II
%%% 

In Figs.\ \ref{Fig4}\figtc~and \figtd, we show the measured qubit linewidth as a function of $\EM$ at $\detuning = 0$ and as a function of $\detuning$ at $\EM = 0$, respectively. Both measurements are performed with constant drive power with an estimated power broadening of $\sqrt{\beta P_\text{s}} \simeq 32$ MHz. 
As described in the Supplementary Information\cite{Supplementary} we numerically model the electric noise-induced dephasing and, by fitting to the data as in Figs.\ \ref{Fig4} \figtc-\figtd, obtain estimates of $\sigma_\detuning / h \simeq 660$ MHz and $\sigma_{\EM} / h\simeq 566$ MHz for the noise amplitudes. 
The decoherence has a minimum at the sweet spot $\delta = \EM = 0$ GHz, however, the rate at which decoherence increases when deviating from the sweet spot is smaller as a function of $\EM$ than $\detuning$. 
These results imply that while the largest contribution to dephasing is dipolar (long-distance) noise, the quadrupolar (short-distance) noise has a comparable magnitude. 
By operating the qubit in the $\EM$ subspace, the qubit experiences lower decoherence than operating in the $\detuning$ subspace since the noise in $\EM$ is smaller than in $\detuning$.
On the other hand, as the ratio $\sigma_{\EM} / \sigma_\detuning \simeq 0.86$ is close to unity, we determine that a substantial contribution to the dephasing originates from noise sources that reside near the TQD.

%%%
%%% PARAGRAPH: CONCLUSIONS
%%%

In conclusion, we have demonstrated strong coupling of a microwave photon to the quadrupole moment of an electron in a triple quantum dot with a coupling strength $g / 2\pi  \simeq 150$ MHz, decoherence rate $\gamma_2 / 2\pi \simeq 32$ MHz and resonator photon decay rate $\kappa / 2\pi  \simeq 18$ MHz. The quadrupole qubit energy is determined by the dipolar and quadrupolar detuning parameters, of which the former is more sensitive to distant voltage fluctuations. We have measured the qubit coherence while detuning from the sweet spot in either of the two parameters and obtained information about the spatial distribution of charge noise in the sample. The observation that electric noise has a substantial contribution to the quadrupolar detuning fluctuations implies that noise sources in close proximity to the quantum dots are relevant. Our results demonstrate that the quadrupolar subspace can provide some protection from decoherence. They provide a promising path towards qubit realizations that store quantum information in quadrupolar states. 

%%%
%%% ACKNOWLEDGEMENTS
%%%

\acknowledgments
We thank Christian Kraglund Andersen and Michele Collodo for useful discussions, and David van Woerkom for his contribution to the sample fabrication.
This work was supported by the Swiss National Science Foundation through the National Center of Competence in Research (NCCR) Quantum Science and Technology. SNC and MF acknowledge support by the Vannevar Bush Faculty Fellowship program sponsored by the Basic Research Office of the Assistant Secretary of Defense for Research and Engineering and funded by the Office of Naval Research through Grant No. N00014-15-1-0029. MR and GB acknowledge funding from ARO through Grant No. W911NF-15-1-0149 and the DFG through SFB 767. MF and JCAU acknowledge support by ARO (W911NF-17-1-0274). The views and conclusions contained herein are those of the authors and should not be interpreted as necessarily representing the official policies or endorsements, either expressed or implied, of the Army Research Office (ARO) or the U.S. Government. The U.S. Government is authorized to reproduce and distribute reprints for Governmental purposes, notwithstanding any copyright notation thereon.

\bibliography{references}

\end{document}